\documentclass[12 pt]{article}

\usepackage[a4paper,bindingoffset=0.2in,%
            left=1in,right=1in,top=1in,bottom=1in,%
            footskip=.25in]{geometry}
\usepackage{authblk}
\usepackage{amssymb}
\usepackage[figuresright]{rotating}
\usepackage[T1]{fontenc}
\usepackage{amsfonts}
\usepackage{dcolumn}
\usepackage{amssymb}
\usepackage{amsmath}
\usepackage{graphicx}
\usepackage{supertabular}
\usepackage{bm}
\usepackage{dsfont}
\usepackage{braket}
\usepackage{hyperref}
\usepackage{color,soul}
\usepackage[english]{babel}

\title{\textbf{Clean Os(0001) electronic surface states: a first-principle fully relativistic investigation}}

\author[1]{Andrea Urru}
\author[1,2]{Andrea Dal Corso}
\affil[1]{International School for Advanced Studies (SISSA),
Via Bonomea 265, 34136 Trieste (Italy).}
\affil[2]{DEMOCRITOS IOM-CNR Trieste (Italy).}

\date{}

\begin{document}
\maketitle

\begin{abstract}
We analyze the electronic structure of the Os(0001) surface by means of
first-principle calculations based on Fully Relativistic (FR) Density
Functional Theory (DFT) and a Projector Augmented-Wave (PAW) approach.
We investigate surface states and resonances analyzing their spin-orbit
induced energy splitting and their spin polarization. The results are
compared with previously studied surfaces Ir(111), Pt(111), and Au(111).
We do not find any surface state in the gap similar to the L-gap of the
(111) fcc surfaces, but find Rashba split resonances that cross the
Fermi level and, as in the recently studied Ir(111) surface, have a
characteristic downward dispersion.
Moreover, for some selected surface states we study the spin
polarization with respect to ${\bf k}_{\parallel}$, the wave-vector
parallel to the surface. In some cases, such as the Rashba split
resonances, the spin polarization shows a smooth behavior with slow
rotations, in others the rotation is faster, due to mixing and
anti-crossing of the states.
\vskip 1.0truecm
\end{abstract}

\newpage
\section{Introduction}

Surfaces can host electronic states localized on the last few layers and  the surface electronic structure is a key ingredient to 
predict many properties of materials. 
Pure surface states are usually found in the gaps of the projected band structure (PBS)\cite{zangwill}, while resonances can be present also within the PBS.
Due to surface states, surfaces can have properties totally different from the bulk, as found e.g. in topological insulators \cite{topological}. 
Moreover, surfaces break inversion symmetry so that even time-reversal invariant nonmagnetic materials 
might have surface states with non-zero spin polarization. 
These states might be practically exploitable, for instance in spintronics applications, so it is important to map them.
The spin polarization of surface states, together with their energy dispersion as a function of ${\bf k}_{\parallel}$, the 
wave-vector parallel to the surface, have been studied in several materials, both theoretically (\cite{au111_first}-\cite{ir111}), mainly 
by means of Density Functional Theory (DFT), and experimentally (\cite{au111_first}, \cite{spin_au}, \cite{L_gap_PES_first}, \cite{L_gap_PES}), 
using photoelectron spectroscopy (PES) \cite{PES}, angular- and spin-resolved.
The L-gap surface states of Au(111) (\cite{au111_first}-\cite{dirac}, \cite{L_gap_PES_first}, \cite{L_gap_PES}) are a paradigmatic example. 
Their main feature is a split parabolic energy dispersion, which can be interpreted as an effect of spin-orbit 
coupling and explained by the Rashba model \cite{rashba}. The latter suggests that the average direction of the electron spin
is perpendicular to the wavevector ${\bf k}_{\parallel}$ and parallel to the surface, with opposite directions in the two 
energy paraboloids, a property that has been proven by experiments \cite{spin_au}, although in real materials the
presence of the underlying atomic layers can give rise to a small spin component orthogonal to the surface.

Similar states have been studied in Ir(111) and Pt(111) surfaces (\cite{KKR}, \cite{ir111}, \cite{H_Pt}-\cite{AlBr3_Ir111}).
Passing from Au to Pt and to Ir their behavior changes: in Au(111) they are found in a PBS gap (the so called L-gap) and show a positive 
curvature dispersion, while in Ir(111) they hybridize with bulk states  and have a characteristic 
negative curvature \cite{Graphene_Ir111}. In Pt(111) their nature turns out to be 
trickier to characterize. At variance with Au(111) they are empty and resemble the Au(111) states away from $\bar{\Gamma}$, 
but close to $\bar{\Gamma}$ they are very close to bulk states and the predicted 
hybridization is quite sensitive to the technical details of the calculation \cite{KKR}, \cite{ir111}.

No information is instead available for Os(0001), a surface very similar to the $(111)$ surface of Ir, Pt, and Au. 
This surface could have states similar to the L-gap surface states with inverted dispersion as in Ir(111) but a-priori 
one cannot exclude the presence of empty surface states in a gap analogous to the L-gap. An obvious difference between Os(0001) and the other
surfaces is the position of the Fermi level, due to the lower atomic number, while more subtle differences could be due to the hexagonal 
close-packed (hcp) structure that on the third layer differs from the face-centered-cubic (fcc) surfaces.

In this paper, we present a first principle study of the Os(0001) band structure and characterize its main surface 
states, including the Rashba split states. We find that they have an inverted dispersion as in Ir(111), and here they cross the 
Fermi level.
For some selected states we study in detail the spin polarization. We first use symmetry considerations to determine its direction in different 
high symmetry lines and points of the Brillouin Zone (BZ) then we follow, for the most interesting surface states, the direction of the spin polarization as a 
function of $\bm{k}_{\parallel}$.
This spin polarization is potentially measurable in spin-resolved ARPES experiments.

The work is organized as follows: in Section 2 we present the methods and the computational parameters. In Section 3 we 
present the Os(0001) band structure and compare the main surface states with those of the other surfaces mentioned above.
In Section 4 we discuss the magnetization density and the spin polarization of selected states. Finally, Section 5 contains our conclusions.

\section{Method}

\begin{figure}
\centering
\includegraphics[width=0.8\textwidth,angle=0]{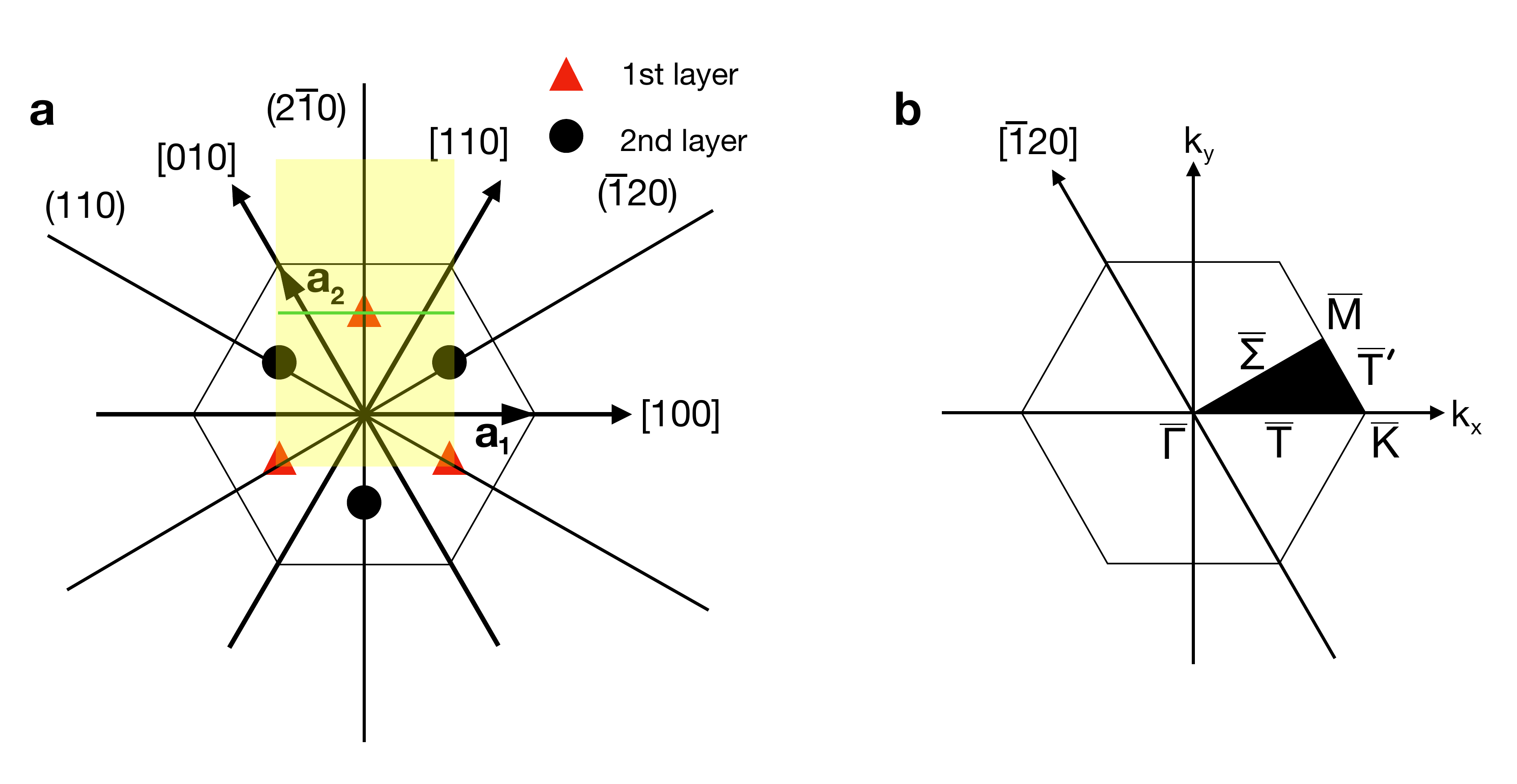}
\caption{(a) Positions of the atoms in the first two atomic layers of the Os(0001) slab. Arrows and solid lines indicate the $C_2$ rotation axes and the mirror planes, respectively.
(b) Surface Brillouin Zone of Os(0001) (in black the Irreducible Brillouin Zone (IBZ)) and the path used to plot the electronic band structure. The $[\bar{1}20]$ axis is the 
two-fold rotation axis of the small group of $\bm{k_{\parallel}}$ along $\bar{T}'$.}
\label{f1}
\end{figure}

First-principle calculations were performed by means of DFT (\cite{HK},\cite{KS}) within the 
Local Density Approximation (LDA), with the Perdew and Zunger's \cite{PZ} parameterization of the exchange and
correlation energy. The Quantum ESPRESSO (\cite{QE}, \cite{QE_2}) and thermo\_pw \cite{thermo_pw} packages are used. 
Spin-orbit coupling effects are included by using the Fully Relativistic (FR) PAW  
method \cite{FR_PAW}, with 5$d$ and 6$s$ valence electrons and 5$s$ and 5$p$ semicore states (Pseudopotential
Os.rel-pz-spn-kjpaw$\_$psl.1.0.0.UPF from pslibrary.1.0.0 \cite{pslibrary}), while Scalar Relativistic
(SR) calculations are performed with the PAW Pseudopotential Os.pz-spn-kjpaw$\_$psl.1.0.0.UPF from 
pslibrary.1.0.0.
Calculations on the bulk system, were performed with an 
hcp structure at the theoretical LDA lattice constants: $a=5.135$ a.u., $c=8.047$ a.u. ($c/a = 1.567$), which
are respectively 0.6\% and 1.4\% smaller than experiment \cite{COD} ($a_{exp} = 5.169$ a.u., $c_{exp} = 8.162$ a.u.). 
The surface has been simulated by a 24-layer slab perpendicular to the (0001) direction separated by a vacuum space of $34$ 
a.u.. The slab crystal structure has been obtained from the bulk, with a further relaxation along 
the [0001] direction. The relaxations are quite small, for instance the distance between the first two layers increases of 
about $0.1 \%$. The pseudo wavefunctions are expanded in a plane waves basis set with a kinetic energy cut-off of 60 Ry, 
while the charge density has a cut-off of 360 Ry. 
BZ integrations were performed using a shifted uniform Monkhorst-Pack \cite{k_grid} $\bf{k}$-point mesh of $12 \times 
12 \times 1$ points for the slab and $12 \times 12 \times 8$ points for the bulk. 
The presence of a Fermi surface has been dealt with by the Methfessel-Paxton method \cite{MP} with a smearing parameter $\sigma=0.02$ Ry. 
With these parameters the total energy is converged within $10^{-3}$ Ry and crystal parameters within $10^{-3}$ $\AA$.
In Fig. \ref{f1}a we show the first two atomic layers of the slab. The slab has a $D_{3d}$ point group. 
In particular, the $z$ axis, normal to the surface, is a $\bar{3}$ rotoinversion axis, while the axes $[100]$, $[110]$, and $[010]$ 
in Fig. \ref{f1}a are two-fold rotation axes. There are also three mirror planes, $(\bar{1}20)$, $(2\bar{1}0)$, and $(110)$ shown in Fig. \ref{f1}a.
The electronic band structure was calculated along the path $\bar{\Gamma}-\bar{K}-\bar{M}-\bar{\Gamma}$ 
(that is along the $\bar{T}$, $\bar{T}'$, and $\bar{\Sigma}$ high-symmmetry lines) of the 
Surface Brillouin Zone (SBZ), shown in Fig. \ref{f1}b.

\begin{figure}[h]
\centering
\includegraphics[width=0.8\textwidth,angle=0]{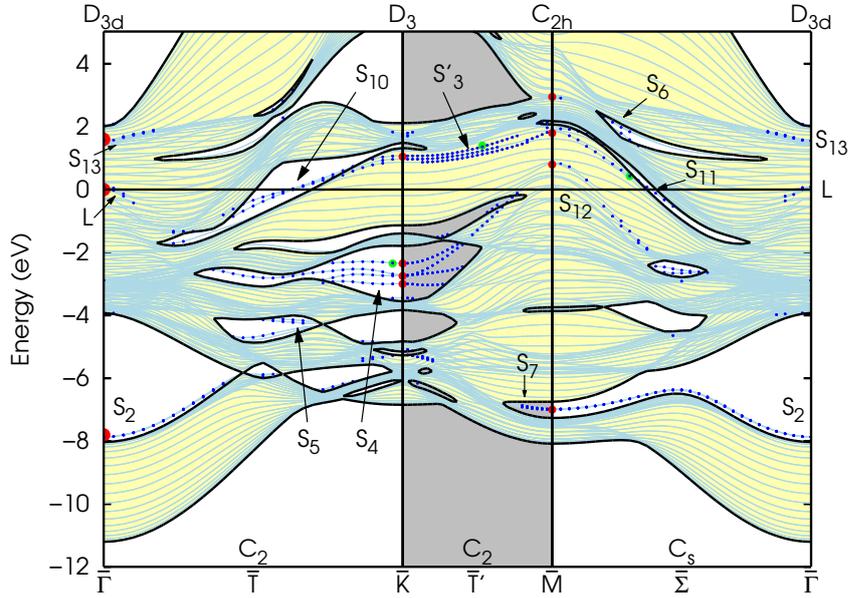}
\caption{(Color online) LDA FR-PAW surface band structure of Os(0001). The yellow region is 
the PBS, the light blue lines are the slab electronic states and the blue dots 
indicate surface states or resonances, defined as those having a charge density greater than 0.5 on the last 
two atomic layers of both surfaces. Energies are measured with respect to the Fermi energy, and the energy maximum 
in the figure is the computed work function (5.03 eV).}
\label{f2}
\end{figure}

\newgeometry{left=0.4in,right=0.4in,top=1in,bottom=1in}
\begin{table*}
\centering
\begin{tabular}{lccccccc}
\hline
\hline
Surface   & ${\bf k}_{\parallel}$ & $\varepsilon$ (eV)& $\varepsilon$ (eV)&$\varepsilon$ (eV)&$\varepsilon$ (eV)&Small group&Symmetry \\
State   & & Os(0001)  & Ir(111) & Pt(111) & Au(111) & of ${\bf k}_{\parallel}$ &  \\
\hline
L          & $\bar \Gamma$ & $0.08$ & $-0.31$ & $0.1$ & $-0.5$ & $D_{3d}\ (C_{3v})$ & $\Gamma_4^-$, $\Gamma_4^+$ $(\Gamma_4)$ \\
$S_{2}$    & $\bar \Gamma$ & $-7.87$ & $-8.0$ & $-7.4$ & $-7.6$ &  & $\Gamma_4^+$, $\Gamma_4^-\ (\Gamma_4)$ \\
$S_{13}$   & $\bar \Gamma$ & $1.55$ & --- & --- & --- & & $\Gamma_4^-$, $\Gamma_4^+$ $(\Gamma_4)$ \\ 
$S'_{3a}$   & $\bar K$      & $1.07$ & --- & --- & --- & $D_3\ (C_3)$  & $\Gamma_5\oplus\Gamma_6\ (2\Gamma_6)$ \\
$S'_{3b}$   & $\bar K$      & $0.97$ & ---  & --- & --- &  & $\Gamma_4\ (\Gamma_4\oplus\Gamma_5)$ \\
$S_{4a}$   & $\bar K$      & $-2.35$ & $-2.7$ & $-2.8$ & $-3.7$ &  & $\Gamma_4\ (\Gamma_4\oplus\Gamma_5)$ \\
$S_{4b}$   & $\bar K$      & $-2.72$ & $-3.1$ & $-3.1$ & $-4.0$ &  & $\Gamma_5\oplus\Gamma_6\ (2\Gamma_6)$ \\
$S_{4c}$   & $\bar K$      & $-2.74$ & $-3.1$ & $-3.1$ & $-4.0$ &  & $\Gamma_4\ (\Gamma_4\oplus\Gamma_5)$ \\
$S_{4d}$   & $\bar K$      & $-2.99$ & $-3.5$ & $-3.7$ & $-4.7$ &  & $\Gamma_4\ (\Gamma_4\oplus\Gamma_5)$ \\
$S_{6}$    & $\bar M$      & $2.93$ & $1.6$  & $0.6$ & --- & $C_{2h}\ (C_s)$ & $\Gamma_3^-\oplus\Gamma_4^- (\Gamma_3\oplus\Gamma_4)$ \\
$S_{11}$    & $\bar M$      & $1.90$ & --- & --- & --- &  & $\Gamma_3^+\oplus\Gamma_4^+$, $\Gamma_3^-\oplus\Gamma_4^-\ (\Gamma_3\oplus\Gamma_4)$ \\
$S_{12}$    & $\bar M$      & $0.86$ & --- & --- & --- & & $\Gamma_3^-\oplus\Gamma_4^- (\Gamma_3\oplus\Gamma_4)$ \\
$S_{7}$    & $\bar M$      & $-7.00$ & $-6.7$ & $-6.3$ & $-6.6$ &  & $\Gamma_3^+\oplus\Gamma_4^+$, $\Gamma_3^-\oplus\Gamma_4^-\ (\Gamma_3\oplus\Gamma_4)$ \\
$S_{10}$   & $0.6\ \bar K$   & $-0.24$ & $-0.8$ & $-1.2$ & --- & $C_2\ (C_1)$ & $\Gamma_3\oplus\Gamma_4$ \\
$S_{12}$   & $0.6\ \bar M$   & $-1.90$ & $-2.6$ & $-2.6$ & --- & $C_s\ (C_s)$ & $\Gamma_3\oplus\Gamma_4$ \\
\hline
\hline
\end{tabular}
\caption{Energy and symmetry properties of the surface states discussed
in the paper for the Os(0001), Ir(111), Pt(111), and Au(111) surfaces. The reported
symmetry refers to the slab. In parenthesis, the symmetry relevant for the surface.
}
\label{t1}
\end{table*}

\newgeometry{left=1in,right=1in,top=1in,bottom=1in}

The small point group of $\bf{k}$ is indicated in the band structure in Fig. \ref{f2}, both for the high symmetry points 
($\bar{\Gamma}$, $\bar{K}$, and $\bar{M}$) and for the high symmetry lines ($\bar{T}$, $\bar{T}'$, and 
$\bar{\Sigma}$). At $\bar{\Gamma}$, $\bar{K}$, and $\bar{M}$ it is $D_{3d}$, $D_{3}$, and $C_{2h}$, respectively. Along the high symmetry lines $\bar{T}$, $\bar{T}'$, and 
$\bar{\Sigma}$ it is $C_2$, $C_2$, and $C_s$, respectively. Along $\bar{T}$ the rotation axis coincides with the $x$-axis, while along
$\bar{T}'$ the rotation axis is the $[\bar{1}20]$ axis, shown in Fig. \ref{f1}b. Finally,
along $\bar{\Sigma}$ the trace of the mirror plane of $C_s$ is $\bar{\Sigma}$.
The slab has more symmetry elements than the Os(0001) surface, since it has symmetry operations that
exchange its two surfaces. Removing these elements,
the surface point group is $C_{3\text{v}}$, while the small groups of $\bf{k}$ are $C_{3\text{v}}$, $C_3$, and $C_s$
for $\bar{\Gamma}$, $\bar{K}$, and $\bar{M}$ respectively and $C_1$, $C_1$, and $C_s$ along $\bar{T}$, 
$\bar{T}'$, and $\bar{\Sigma}$.

\section{Results}

In this section we analyze the FR band structure of Os(0001), shown in Fig. \ref{f2}. We 
characterize the main surface states and compare with Au(111), Pt(111), and Ir(111) (Refs. \cite{au111}, \cite{ir111}).
A list of the main surface states, their energy and symmetry, is given in Table \ref{t1}. 
We start our analysis from the $\bar{\Gamma}$ point, where we find two main gaps in the PBS. Taking the energy zero at the 
Fermi energy, the highest starts at $2$ eV and is similar to the L-gap of the fcc surfaces. It is located higher in 
energy with respect to Au(111), Pt(111), and Ir(111), due to the lower number of electrons per cell.  
At $\bar{\Gamma}$ we find a second gap, between $-8 $ eV $< E < -4 $ eV, approximately
at the same energy as in Pt(111) and Ir(111), although a bit wider and extending up to half of the $\bar{T}$ line and along the
whole $\bar{\Sigma}$ line.
As in Ir(111), and at variance with Au(111) and Pt(111), no surface states are found in the L-gap. Rashba-split
surface states similar to the L-gap states of Au(111) are found around 
$\bar{\Gamma}$ near the Fermi energy and, as in Ir(111), show a characteristic negative dispersion. At $\bar{\Gamma}$ there are two 
groups of two degenerate states, transforming as the $\Gamma_{4+}$ and $\Gamma_{4-}$ representations of the $D_{3d}$ 
group, that extend up to 0.15 $\AA^{-1}$ along $\bar{T}$ and to 0.13 $\AA^{-1}$ along 
$\bar{\Sigma}$. Due to the finiteness of the slab the two surfaces are not perfectly decoupled, so the two groups of states
are not exactly degenerate at $\bar{\Gamma}$, although their energy splitting is very small ($\approx 0.01 $ eV). Neglecting this 
splitting it is possible to fit them with two parabolas as in the Rashba model \cite{rashba}:
\begin{equation}
\centering
E_{\pm} = \frac{\hbar^2}{2 m^*} k_{\parallel}^2 \pm \gamma_{SO} k_{\parallel},
\label{eq1}
\end{equation}
where
\begin{figure}
\centering
\includegraphics[height=0.6\textwidth,angle=-90]{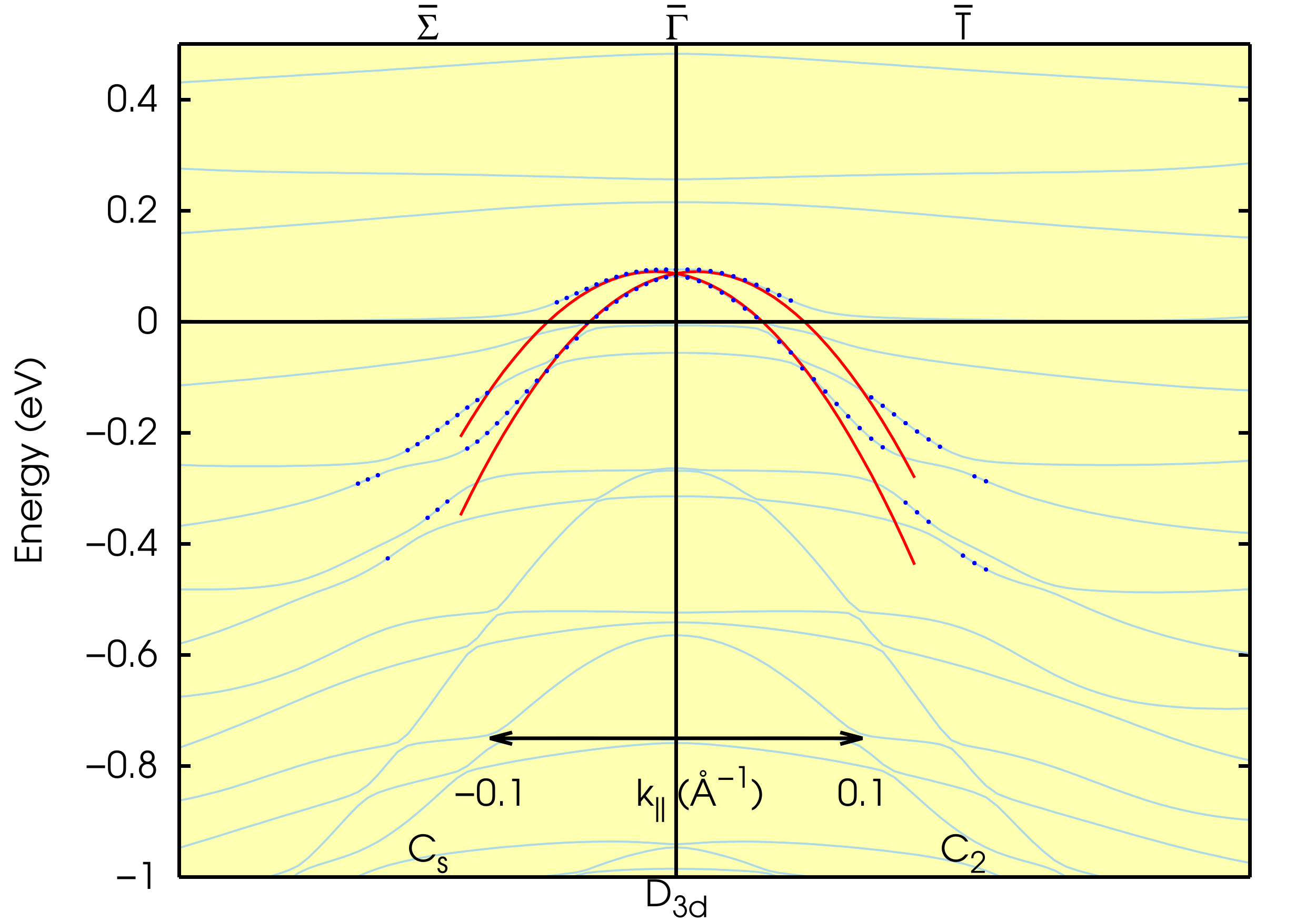}
\caption{Magnification of the FR electronic band structure around $\bar{\Gamma}$. The $L$ states are shown with blue dots. Red lines show the two Rashba 
parabolas that fit the energy dispersion, with the parameters reported in the main text.}
\label{f9}
\end{figure}
\begin{figure}
\centering
\includegraphics[width=0.9\textwidth,angle=0]{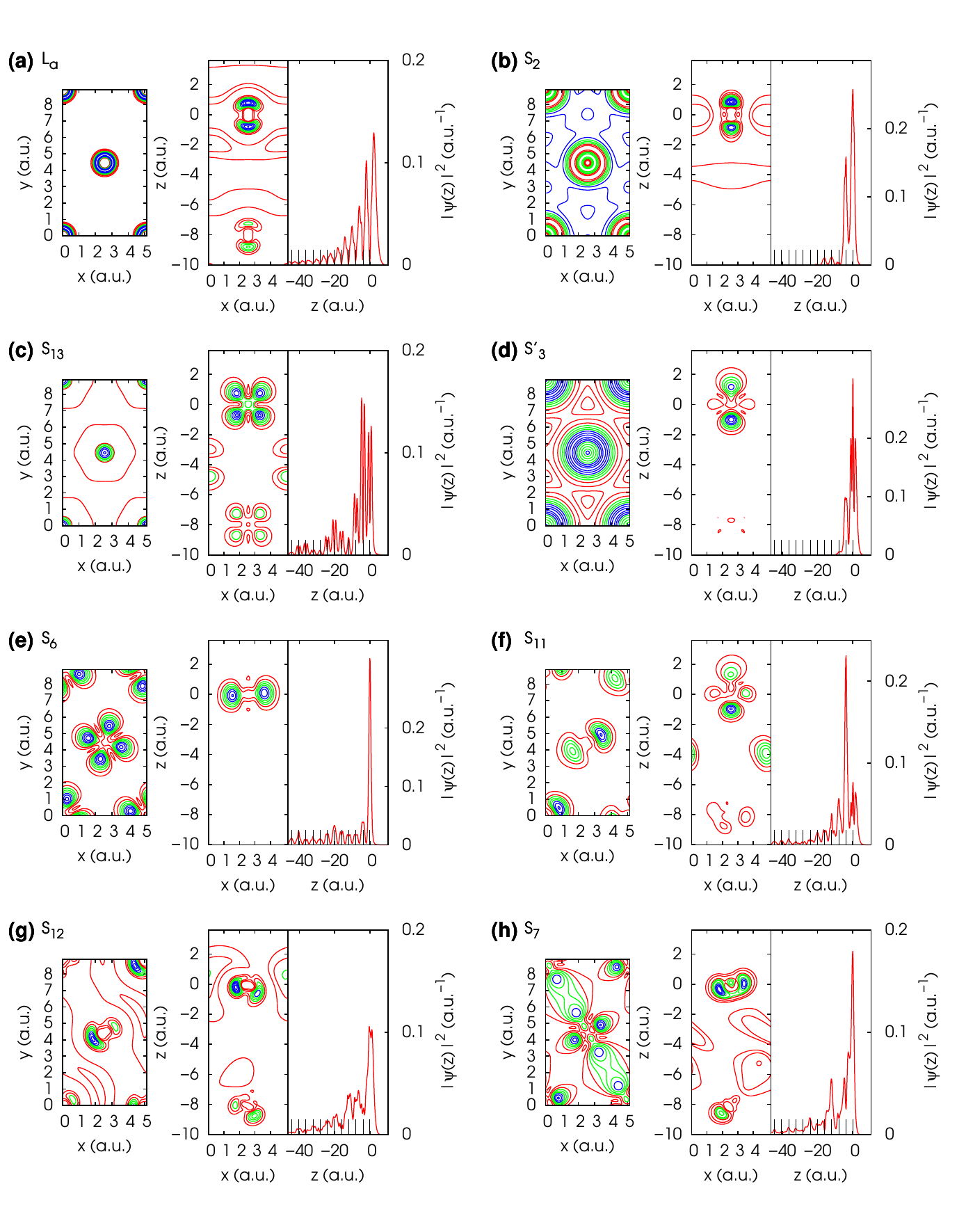}
\caption{Contour plots and planar average of the charge density corresponding to the selected FR surface states indicated with red dots in Fig. \ref{f2}.
The left subplot shows the charge density contour plot in the yellow region in Fig. \ref{f1}a, on the top atomic layer of the slab. The central subplot 
shows the contour plot in a plane perpendicular to the slab, whose trace is the green line
in Fig. \ref{f1}a. The contours are equally spaced and are indicated with different colors (red, green, and blue in increasing order of charge density).
The first three atomic layers are shown. The right subplot shows the planar average of the charge density 
in one half of the slab. The vacuum is on the right; the $x$ tics represent the positions of the atomic layers.}
\label{f3}
\end{figure} 
$k_{\parallel}$ is the modulus of the wave-vector parallel to the surface, $m^*$ is the effective electron mass
and $\gamma_{SO}$ is the spin-orbit coupling parameter.
$\gamma_{SO}$ and $m^*$ are obtained by fitting $E_+ - E_-$ as a function of $k_{\parallel}$
with a straight line and $E_+ + E_-$ with a parabola centered in $\bar{\Gamma}$, respectively. The fit of our data,
shown in Fig. \ref{f9}, gives: 
$\gamma_{SO} = (6.1 \pm 0.1) \times 10^{-9} $ eV cm, which falls between the values found in Au(111) and Ir(111) 
($\gamma_{SO} = 4.5 \times 10^{-9} $ eV cm \cite{au111} and $\gamma_{SO} = 13 \times 10^{-9} $ eV cm \cite{ir111} respectively). The fitted value of
$m^*/m= -0.140 \pm 0.001$, where $m$ is the electron mass, is approximately $40 \%$ lower in modulus than in Au(111) ($0.24$) and Ir(111) 
($-0.22$) \cite{40_layers}. Along $\bar{\Sigma}$ we find $m^*/m= -0.146 \pm 0.002$ and
a $\gamma_{SO}$ parameter equal, within the error bar, to the one obtained along $\bar{T}$.
\begin{figure}
\centering
\includegraphics[width=0.9 \textwidth,angle=0]{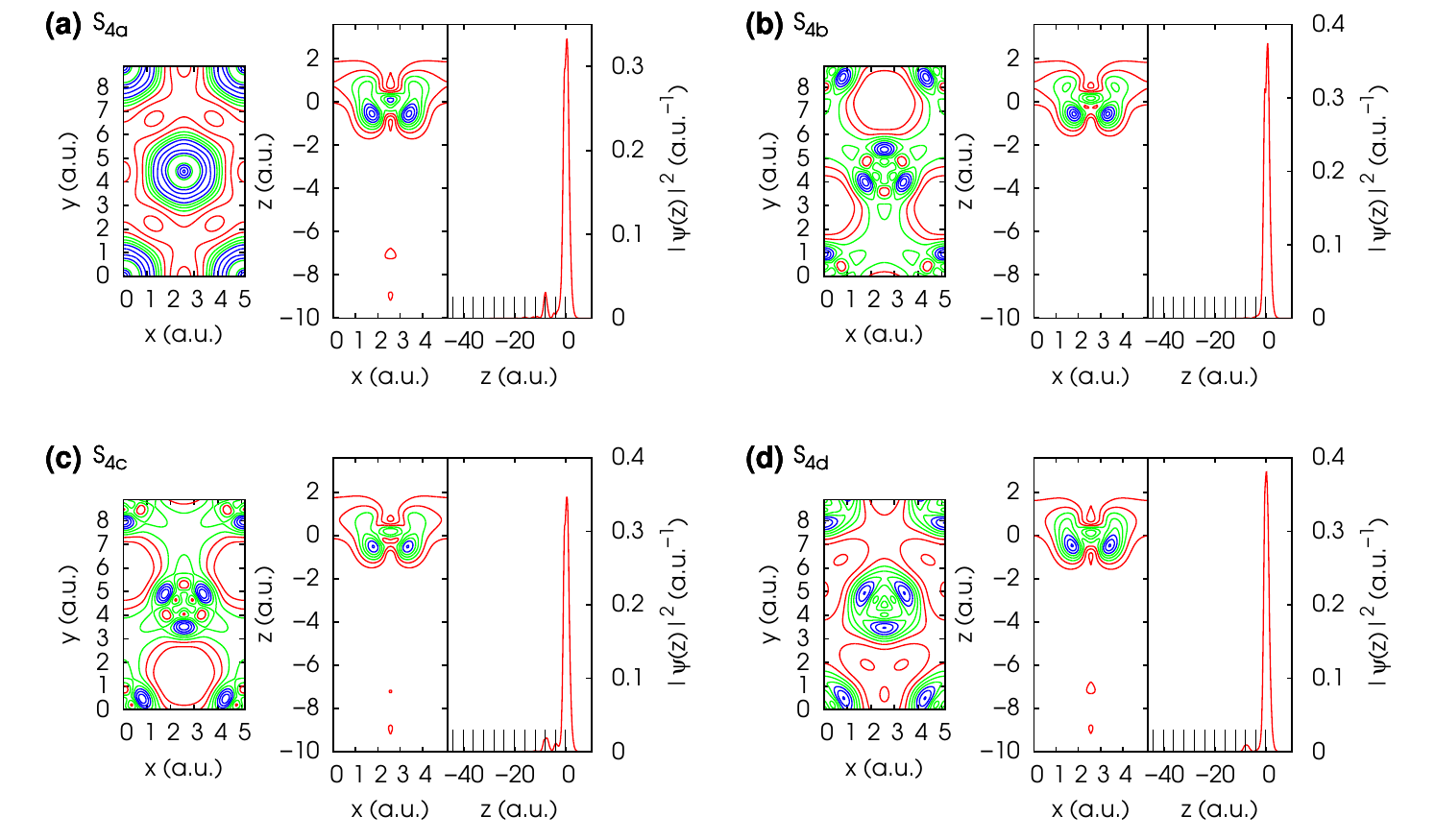}
\caption{Contour plots and planar average of the charge density of the $S_4$ surface states at $\bar{K}$. 
The organization of the subplots is the same as in Fig. \ref{f3}.}
\label{f4}
\end{figure}
In Fig. \ref{f3}a we show the sum of charge densities of the two degenerate states at higher energy that form the $L_a$ band.
The planar average of the charge density is maximum at the surface and decays toward the center of the slab. The contour plots
suggest that it has mainly $s$ character hybridized with some $d$ states, as also confirmed by the projection on atomic wavefunctions. We find very 
small projections on the unoccupied $6p$ states.

At lower energies, at the $\bar{\Gamma}$ point, we find a group of two two-fold degenerate states within a PBS gap, 
similar to the $S_2$ states previously studied in Au(111), Pt(111), and Ir(111) \cite{PES_DFT_Ir111}. 
At $\bar{\Gamma}$, the $S_2$ states have $\Gamma_{4+}$ and $\Gamma_{4-}$ symmetry, as the $L$ states. Their
energy difference is linear in $k_{\parallel}$, as for Rashba split states, 
although with a smaller value of the spin-orbit parameter. A fit performed as above gives: $\gamma_{SO} = (0.288 \pm 0.003) \times
10^{-9} $ eV cm and $m^*/m = 0.732 \pm 0.005$, with identical values, within the error bar, along $\bar{T}$ and $\bar{\Sigma}$.
In Fig. \ref{f3}b we show the charge density for the  $S_{2a}$ states, those with higher energy at $\bar{\Gamma}$. 
The states are localized in the last two atomic layers.
 
Finally, in $\bar{\Gamma}$ there is another group of states, called $S_{13}$ in Fig. \ref{f2}, that was not discussed before. 
They appear in the relaxed surface and, at $\bar{\Gamma}$, they have
symmetry $\Gamma_{4+}$ and $\Gamma_{4-}$. 
As the L states, they are resonances, as shown by the planar average of the charge density in 
Fig. \ref{f3}c. The contour plots, together with the analysis of the projection
of the states on atomic orbitals, show that the $S_{13}$ states have mainly $d$ character, with main projections on the second and first
atomic layers.

$L$, $S_{2}$, and $S_{13}$ states extend also along a portion of the $\bar{T}$ line, where they transform as the $\Gamma_3 \oplus \Gamma_4$ 
representations of $C_2$. Along $\bar{T}$ we find other PBS gaps as well: the widest 
ones contain also some surface states as $S_{10}$, that cross the Fermi level, and $S_4$, at lower energies. 
A small energy gap, in the central part of the $\bar{T}$ line, contains the $S_5$ states. 
At variance with Au(111), Pt(111), and Ir(111) the $S_{10}$ states are inside a PBS gap, while they were hybridized with the bulk
in the other surfaces.

The main states at the $\bar{K}$ point are $S'_3$ (3 couples of empty states) and $S_4$ (4 couples of occupied states). 
$S'_{3a}$ have symmetry $\Gamma_5 \oplus \Gamma_6$ ($D_3$ group), while $S'_{3b}$ have symmetry $\Gamma_4$ ($D_3$ group). 
The planar average and contour plots of the charge density of $S'_{3a}$ are shown in Fig. \ref{f3}d.
These states are almost entirely localized in the first two layers and derive mainly from $d_{3z^2-r^2}$ states. A comparison
with the $S'_3$ states in Ir(111) shows that they have a similar character, although in Ir(111) they are
in a PBS gap.
\begin{figure}[t]
\centering
\includegraphics[height=0.5\textwidth,angle=270]{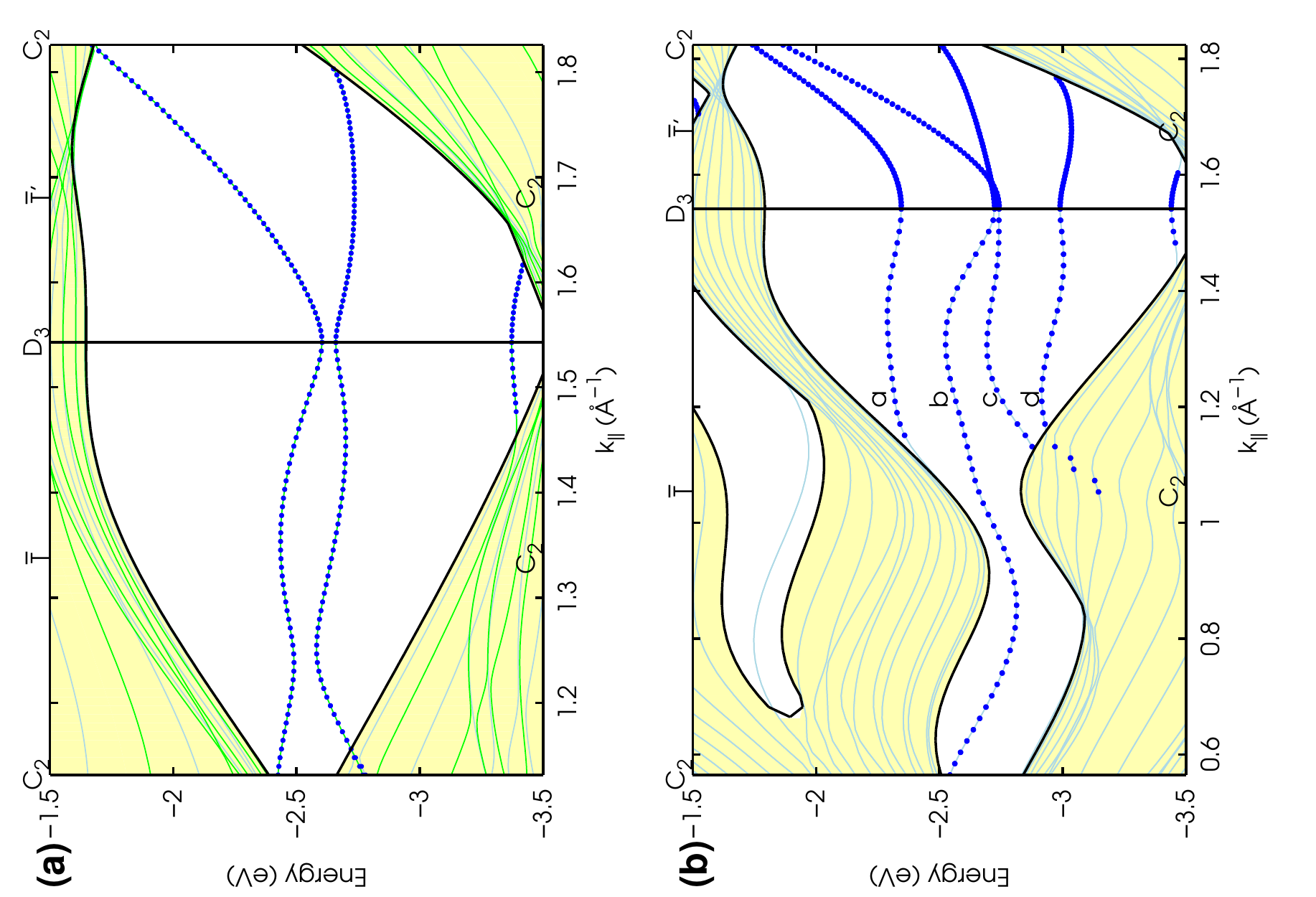}
\caption{Enlarged view of the $S_4$ surface states around $\bar{K}$ in the SR (a) and FR (b) case, respectively.}
\label{f10}
\end{figure}

In the PBS gap located at $-3.8 $ eV $< E < -1.8 $ eV we find the $S_4$ states, whose charge densities are shown in Fig. \ref{f4}.
They are strongly localized on the top atomic layer, with a very small contribution in the third layer for $S_{4a}$, $S_{4c}$,
and $S_{4d}$. To analyze in more detail their nature, in Fig. \ref{f10} we compare the FR $S_4$ states with those found in the SR scheme. 
In the SR case (Fig. \ref{f10}a), the $S_4$ states are two couples of degenerate 
states. Along $\bar{T}$ 
($\bar{T}'$) each couple is made up of an even and an odd state with respect to the $C_2$ rotation about the $x$ ($[\bar{1}20]$) axis.
Their degeneracy is due to the localization of the states in the outermost atomic layers, not to symmetry.
Only at $\bar{K}$, the two states at higher energy are exactly degenerate and transform as the $E$ representation of the $D_{3}$ group, while 
the states at lower energy have $A_1$ and $A_2$ symmetry, respectively. The former projects with similar weights on 
$d_{x^2-y^2}$, $d_{xy}$, $d_{xz}$, and $d_{zy}$ of the first layer, while the latter projects mainly on 
$d_{xy}$ and $d_{x^2-y^2}$ states, with small projections on $d_{xz}$ and $d_{yz}$.

In the FR band structure, along the $\bar{T}$ and $\bar{T}'$ lines all the states have symmetry $\Gamma_3 \oplus \Gamma_4$ of the group $C_2$, consistent
with the product of the SR representation with the $D_{1/2}^{+}$ representation of spin ($A \otimes D_{1/2}^{+} = \Gamma_3 \oplus \Gamma_4$,
$B \otimes D_{1/2}^{+} = \Gamma_3 \oplus \Gamma_4$). At $\bar{K}$ $S_{4a}$, $S_{4c}$, and $S_{4d}$ have symmetry $\Gamma_4$
(group $D_3$), while $S_{4b}$ has symmetry $\Gamma_5 \oplus \Gamma_6$. Multiplying the SR symmetries with $D_{1/2}^{+}$ we have
$A_1 \otimes D_{1/2}^{+} = \Gamma_4$, $A_2 \otimes D_{1/2}^{+} = \Gamma_4$, $E \otimes D_{1/2}^{+} = \Gamma_4 \oplus \Gamma_5 \oplus \Gamma_6$ and,
as a result, the FR $S_4$ states mix both the SR states.
Another consequence of the spin-orbit effects is the anti-crossing of the states $S_{4b}$ and $S_{4c}$ near $\bar{K}$ ($k_{\parallel} = 1.59 \, \AA^{-1}$).

The surface states found in $\bar{K}$ extend also partly along $\bar{T}'$ but do not reach $\bar{M}$. 
Near $\bar{M}$ there are three gaps in the PBS. Two of them, at $2$ eV and $-3.8$ eV, are quite narrow, 
while the third one is between $-7$ eV $< E < -6.5$ eV. At variance with Au(111),
Pt(111), and Ir(111) we do not find the gap that contained
the Dirac-like $S_8$ states studied in Ref. \cite{dirac}, and $S_8$  states are not found 
in Os(0001). Moreover, the gap along $\bar{\Sigma}$ that contains the $S_2$ states extends up to the
$\bar{M}$ point and includes also the $S_7$ surface states.
The other surface states at $\bar{M}$ are, in decreasing order of energy, $S_6$, $S_{11}$, and $S_{12}$.
$S_6$ is empty, has symmetry $\Gamma_{3-} \oplus \Gamma_{4-}$ ($C_{2h}$ group) and is a resonance, as can 
be seen from the planar average of its charge density in Fig. \ref{f3}e.
This state, present in Au(111) (SR) band structure, disappears when spin-orbit coupling is included (see \cite{au111} 
for more details). It is present also in the FR band structure of Pt(111) and Ir(111), but it is not located in a 
PBS gap as in the SR Au(111) surface. The charge density contours of the $S_6$ states in Au(111) are pretty similar 
to those of Os(0001), both in the top atomic layer and perpendicularly to the surface. 
At $\bar{M}$ we find also the states $S_{11}$ and $S_{12}$, that belong to the
representations $\Gamma_{3+} \oplus \Gamma_{4+}$ and $\Gamma_{3-} \oplus \Gamma_{4-}$ of $C_{2h}$ respectively. 
They are both resonances, with the main charge contribution coming from second 
($S_{11}$) or first ($S_{12}$) layer. $S_{11}$ have mainly $d_{3z^2-r^2}$ character, as shown by the contour plot in the 
plane perpendicular to the surface, and smaller components coming from first layer $d_{xy}$ and $d_{x^2-y^2}$ states. 
Even though we gave them the same names as the states present in the same energy region in Ir(111), $S_{11}$ and $S_{12}$ look 
quite different from those of Ref. \cite{ir111}. $S_{12}$ has  
non-negligible projections on many $d$ states of the first six layers of the slab and small projection on $s$ states.
Finally, again at $\bar{M}$, but deeper in energy we have the $S_7$ states, arranged in two
couples of degenerate states, with symmetry $\Gamma_{3+} \oplus \Gamma_{4+}$ and $\Gamma_{3-} \oplus \Gamma_{4-}$ respectively. 
In Fig. \ref{f3}h we report the contour plots and planar average of the charge density of $S_7$, 
which shows relevant contributions on the first, second and fourth layer. The $S_7$ states of Os(0001) are in a PBS gap and
differ somewhat from those of Au(111).
$S_7$ projects on many atomic wavefunctions: the main contributions come from $d$ 
states of the first two layers and $s$ states of the top layer. 

\section{Spin polarization: results and discussion}

\begin{figure}[t!]
\centering
\includegraphics[width=.8\textwidth,angle=0]{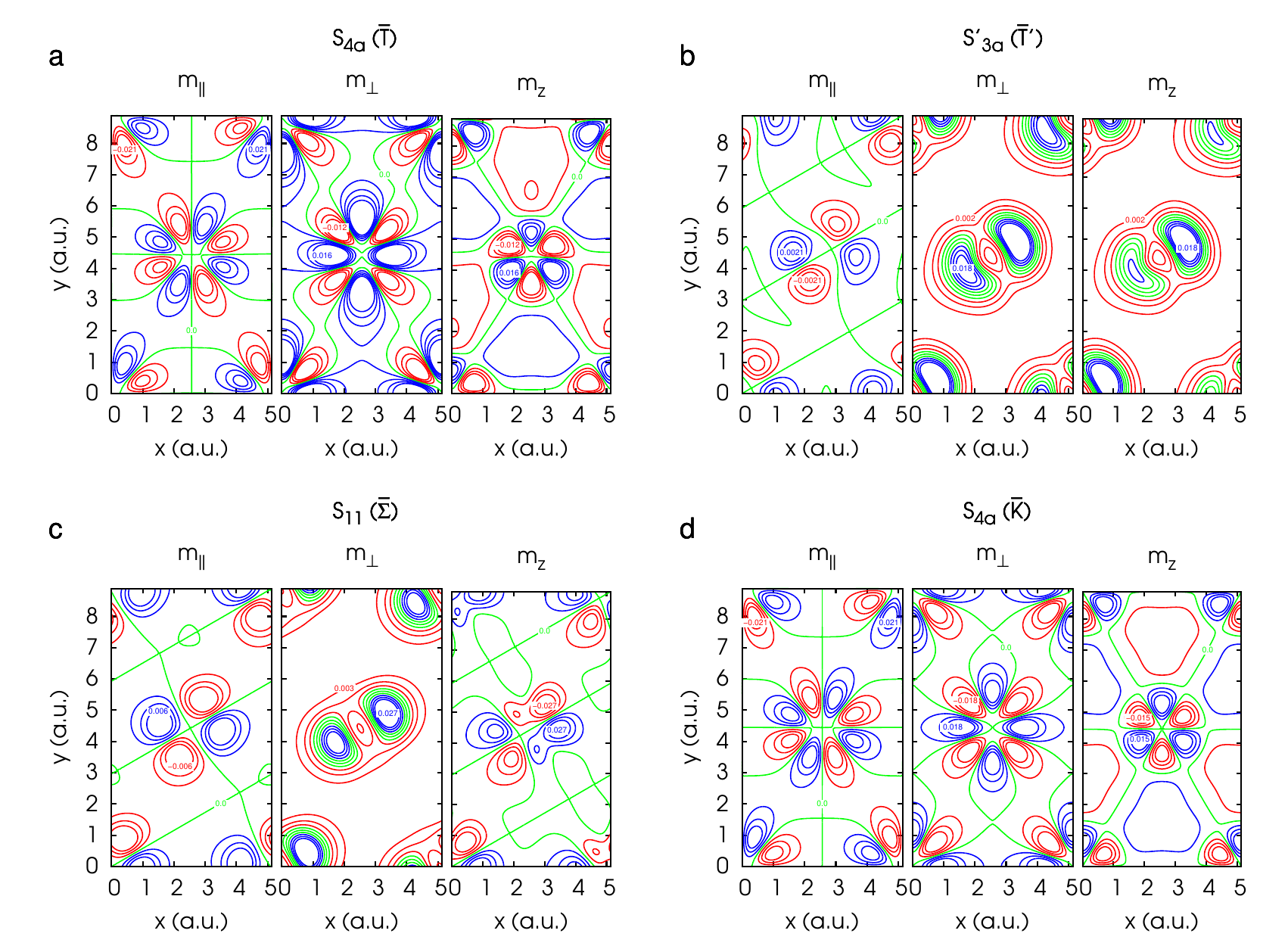}
\caption{Magnetization density contour plots for the 
selected FR surface states, indicated with green dots in Fig. \ref{f2}. The three subplots show, from left to right, the contour plot
of the components parallel and perpendicular to the high symmetry line, and perpendicular to the slab. The contour levels are equally 
spaced and range from a minimum and a maximum value, indicated in the figure.}
\label{f5}
\end{figure}

In this section we discuss the spin polarization of some of the surface states found above. 
The spin polarization can be obtained integrating the planar average of the magnetization density over half slab:
\begin{equation}
\centering
m_{\alpha}=\int_{0}^{L/2} m_{\alpha} (z) \, dz \, ,
\label{eq6}
\end{equation}
where the zero of $z$ is taken at the center of the slab and $L$ is its length along $z$, including vacuum. $m_{\alpha} (z)$ in Eq. \ref{eq6} is
the planar average of the magnetization density $m_{\bm{k}n}^{\alpha} (\bm{r})$ associated to the Bloch state 
$\braket{\bm{r} | \Psi_{\bm{k} n \sigma}}$ and is defined as:
\begin{equation}
\centering
m_{\alpha}(z) = \int_{A} m_{\bm{k}n}^{\alpha} (x,y,z) \, dx \, dy \, ,
\label{eq5}
\end{equation}
where $A$ is the yellow shaded region shown in Fig. \ref{f1}a, and 
\begin{equation} 
\centering
m_{\bm{k}n}^{\alpha} (\bm{r})= \mu_B \sum_{\sigma_1,\sigma_2} \braket{\Psi_{\bm{k} n \sigma_1} | \bm{r}} \sigma^{\sigma_1 \sigma_2}_{\alpha} \braket{\bm{r} | \Psi_{\bm{k} n \sigma_2}},
\label{eq2}
\end{equation}
where $\mu_B$ is the Bohr magneton and $\sigma_{\alpha}$ are the Pauli matrices. The sum over $\sigma_1$ and $\sigma_2$ is over the spin.

Bulk Os and Os(0001) slab have inversion symmetry, and since Os is nonmagnetic and its Hamiltonian is time-reversal invariant, all the 
bands are at least two-fold degenerate. Eq. \ref{eq2} must be generalized accordingly considering the sum of the contributions 
of the degenerate states to the magnetization. The surface breaks inversion symmetry, hence we expect the states to have a 
non-zero average spin polarization when integrating the sum of the two magnetization densities in half slab. 

The crystal possesses other symmetries as well, like rotations ($\mathcal{S}$), possibly together with fractional translations 
($\bm{f}$). They can induce some constraint on the magnetization density, leading in some cases to a vanishing spin polarization. 
If $(\mathcal{S},\bm{f})$ is an operation of the small space group of $\bm{k}$, or if it is when composed with time reversal 
$\mathcal{T}$, the sum of the magnetization densities of degenerate states must obey the following relationship:

\begin{equation}
\centering
\bm{m}(\bm{r}) = \pm \tilde{\mathcal{S}} \, \bm{m}(\mathcal{S}^{-1} \bm{r} - \mathcal{S}^{-1} \bm{f}),
\label{eq3}
\end{equation}
where $\tilde{\mathcal{S}}$ is the proper part of $\mathcal{S}$. The $\pm$ signs of Eq. \ref{eq3} distinguish the operations that 
require $\mathcal{T}$ ($-$ sign) from those that do not require it ($+$ sign) and are due to the fact that $\mathcal{T}$ reverts 
the sign of the magnetization.

Os(0001) space group does not contain any fractional translation, so we have $\bm{f} = \bm{0}$, and in the following we focus on the
small point group of $\bm{k}$.

In addition to its operations one can consider the magnetic 
point group, obtained by multiplying the small point group operations by $\{E, \mathcal{T} \mathcal{I}\}$ ($\mathcal{I}$ is the inversion), 
since $\mathcal{T} \mathcal{I}$ leaves $\bm{k}$ invariant and $\mathcal{I}$ is contained in the slab point group $D_{3d}$. 
The magnetic point groups for the high symmetry lines and points of Os(0001) slab are summed up in the central column of Table \ref{t2}. 
For the surface we must remove from these groups the operations that exchange the two slab surfaces, obtaining the groups listed in the 
right column of Table \ref{t2}.

For a group that contains $\mathcal{T}$, Eq. \ref{eq3} implies, using $\mathcal{S}=\mathds{1}$,
that $\bm{m}(\bm{r}) = -\bm{m}(\bm{r})$, so $\bm{m}(\bm{r}) = \bm{0}$. In Os(0001) this is the case for the small point group of
the time-reversal invariant $\bm{k}$ points $\bar{\Gamma}$ and $\bar{M}$. 

The states at $\bar{T}$, $\bar{T}'$, $\bar{\Sigma}$, and $\bar{K}$, have instead a non-zero magnetization density.
We show it in Fig. \ref{f5} on the plane $xy$ (the yellow shaded area in Fig. \ref{f1}a) for the states highlighted with 
green dots in Fig. \ref{f2} (a state in each line or point).

\begin{table}
\centering
\begin{tabular}{ccc}
\hline
${\bf k}_{\parallel}$ & Slab & Surface \\
\hline
$\bar{\Gamma}$ & $D_{3d} \otimes \{E, \mathcal{T} \}$ & $C_{3v} \otimes \{E, \mathcal{T} \}$\\
$\bar{T}$ & $C_{2h} (C_2)$ & $C_{s} (C_1)$ \\
$\bar{K}$ & $D_{3d} (D_3)$& $C_{3v} (C_3)$\\
$\bar{T}'$ & $C_{2h} (C_2)$& $C_{s} (C_1)$\\
$\bar{M}$ & $C_{2h} \otimes \{E, \mathcal{T} \}$& $C_{s} \otimes \{E, \mathcal{T} \}$\\
$\bar{\Sigma}$ & $C_{2h} (C_s)$& $C_{s} (C_s)$\\
\hline
\end{tabular}
\caption{Small groups of $\bf{k}_{\parallel}$ at high symmetry points and high symmetry lines for the slab (central 
column) and a single surface (right column).}
\label{t2}
\end{table}

\begin{figure}[t]
\centering
\includegraphics[width=.8\textwidth,angle=0]{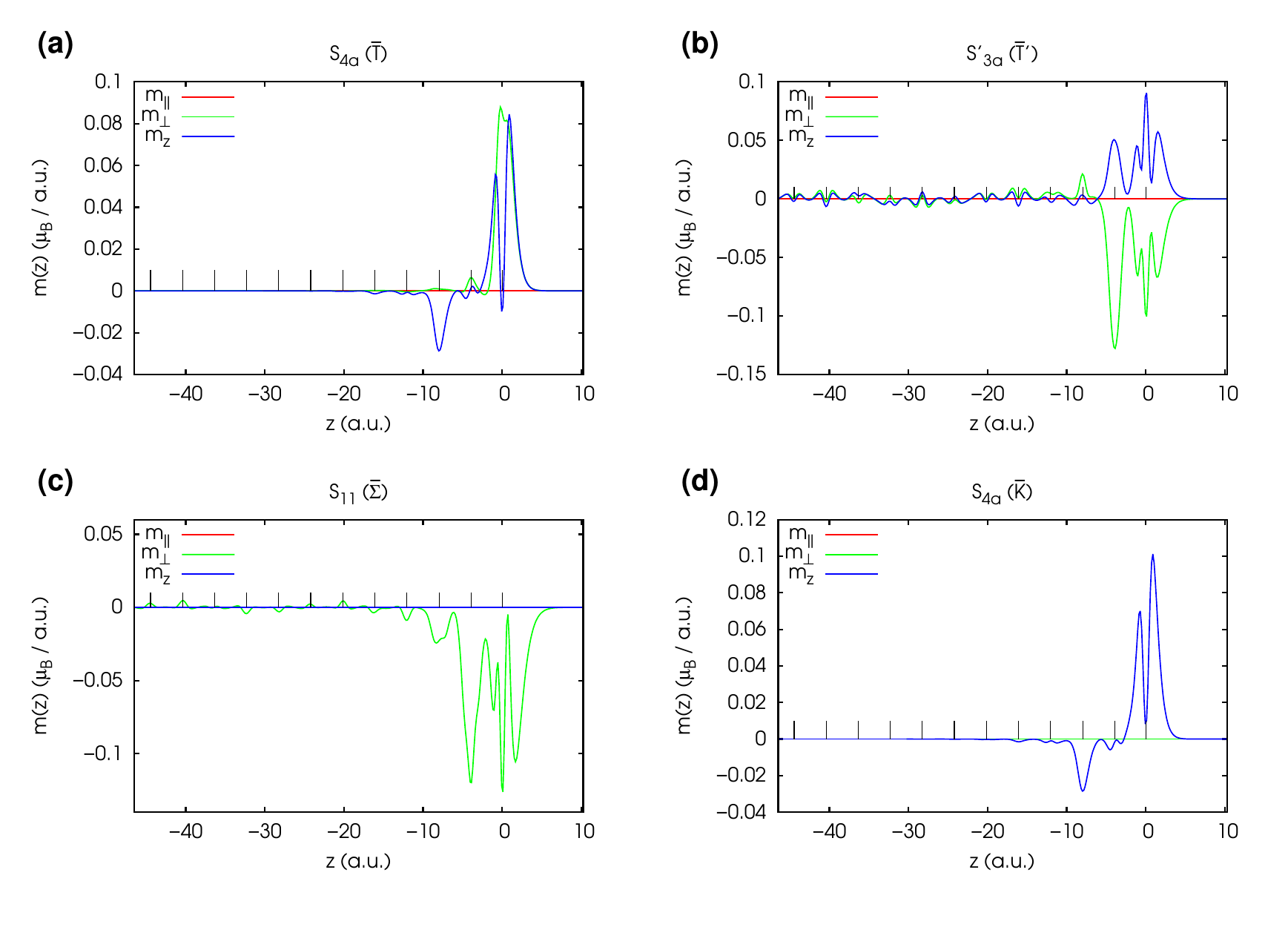}
\caption{Planar average of the magnetization density for the same states as in Fig. \ref{f5}. Only one half of the slab is shown. The vacuum is on the right of each plot,
while the tics in the reference line represent the position of the atomic layers.}
\label{f6}
\end{figure}

Along $\bar{T}$ the small magnetic point group of $\bm{k_{\parallel}}$, $C_{s} (C_1)$, contains 
$\mathcal{T} \mathcal{I} C_{2x} = \mathcal{T} \sigma_y$ ($\sigma_y$ is the mirror plane perpendicular to $x$), which inserted in Eq. \ref{eq3} leads to 
$m_{\parallel}(x,y,z) = -m_{\parallel}(-x,y,z)$, $m_{\perp}(x,y,z) = m_{\perp}(-x,y,z)$, and $m_{z}(x,y,z) = -m_{z}(-x,y,z)$, 
since $\tilde{\mathcal{S}} = C_{2x}$, the two-fold rotation about $x$. As an example we show the state $S_{4a}$ in Fig. \ref{f5}a.

Along $\bar{T}'$ the small magnetic point group of $\bm{k_{\parallel}}$, $C_s(C_1)$, 
has elements $\{E, \mathcal{T} \sigma_{\bar{\Sigma}}\}$, where $\sigma_{\bar{\Sigma}}$ is 
the mirror plane $(\bar{1}20)$. Considering $\mathcal{T} \sigma_{\bar{\Sigma}}$ in 
Eq. \ref{eq3}, we find that $m_{\parallel}$ has opposite signs in the two sides of the mirror $\sigma_{\bar{\Sigma}}$,
while $m_{\perp}$ and $m_z$ have the same sign, as confirmed by the state $S'_{3a}$ (Fig. \ref{f5}b).

Along $\bar{\Sigma}$, the small magnetic point group of $\bm{k_{\parallel}}$, $C_s(C_s)$,
contains only the operations $E$ and $\sigma_{\bar{\Sigma}}$ and not $\mathcal{T}$ or operations that require $\mathcal{T}$. 
Considering the operation $\sigma_{\bar{\Sigma}}$ in Eq. \ref{eq3} one finds that $m_{\parallel}$ and $m_z$ change sign in the two 
sides of the mirror $\sigma_{\bar{\Sigma}}$, while $m_{\perp}$ does not, as shown by the state $S_{11}$ in Fig. \ref{f5}c.
 
Finally, at the $\bar{K}$ point the small magnetic point group of $\bm{k_{\parallel}}$ is $C_{3\text{v}}(C_3)$, 
with operations $\{E$, $C_3$, $C_3^2$, $\mathcal{T} \sigma_y$, $\mathcal{T} \sigma_{\bar{\Sigma}}$, 
$\mathcal{T} \sigma_{\bar{\Sigma}'} \}$, where $\sigma_{\bar{\Sigma}'}$ is the mirror plane $(110)$.
Among these, $\mathcal{T} \sigma_y$ leads to $m_{\parallel}(x,y,z) = - m_{\parallel}(-x,y,z)$, 
$m_{\perp}(x,y,z) = m_{\perp}(-x,y,z)$, $m_{z}(x,y,z) = m_{z}(-x,y,z)$, as along $\bar{T}$. $C_3$ and $C_3^2$ operations 
lead to a three-fold rotational symmetry for the $m_z$ component, whereas $\mathcal{T} \sigma_{\bar{\Sigma}}$ and 
$\mathcal{T} \sigma_{\bar{\Sigma}'}$ lead to the mirror planes $(\bar{1}20)$ and $(110)$ for the $m_z$ component, while they impose more 
complex constraints on $m_{\parallel}$ and $m_{\perp}$. As an example we show the state $S_{4a}$ in Fig. \ref{f5}d. 

In Fig. \ref{f6} we show, for the same states as in Fig. \ref{f5}, the planar averages of the components of the magnetization density. 
The $S_{4a}$ states at $\bar{T}$ and $\bar{K}$ show a magnetization that vanishes after five layers below the surface, while
$S'_{3a}$ and $S_{11}$ have some non negligible contribution also in the center of the slab. As a consequence of the symmetries 
discussed above, all states have a vanishing component $m_{\parallel}(z)$ parallel to the high symmetry line: as a result the spin
polarization lies in a plane perpendicular to the high symmetry line. The states along $\bar{\Sigma}$ have also a vanishing
$z$ component, so their spin polarization has a fixed direction, parallel to the surface. Moreover, since $\bar{K}$ belongs both to 
$\bar{T}$ and $\bar{T}'$, both the components parallel to $\bar{T}$ and $\bar{T}'$ must vanish, so the states at $\bar{K}$ have only a 
$z$ component. 

\begin{figure}[t]
\centering
\includegraphics[height=0.8\textwidth,angle=-90]{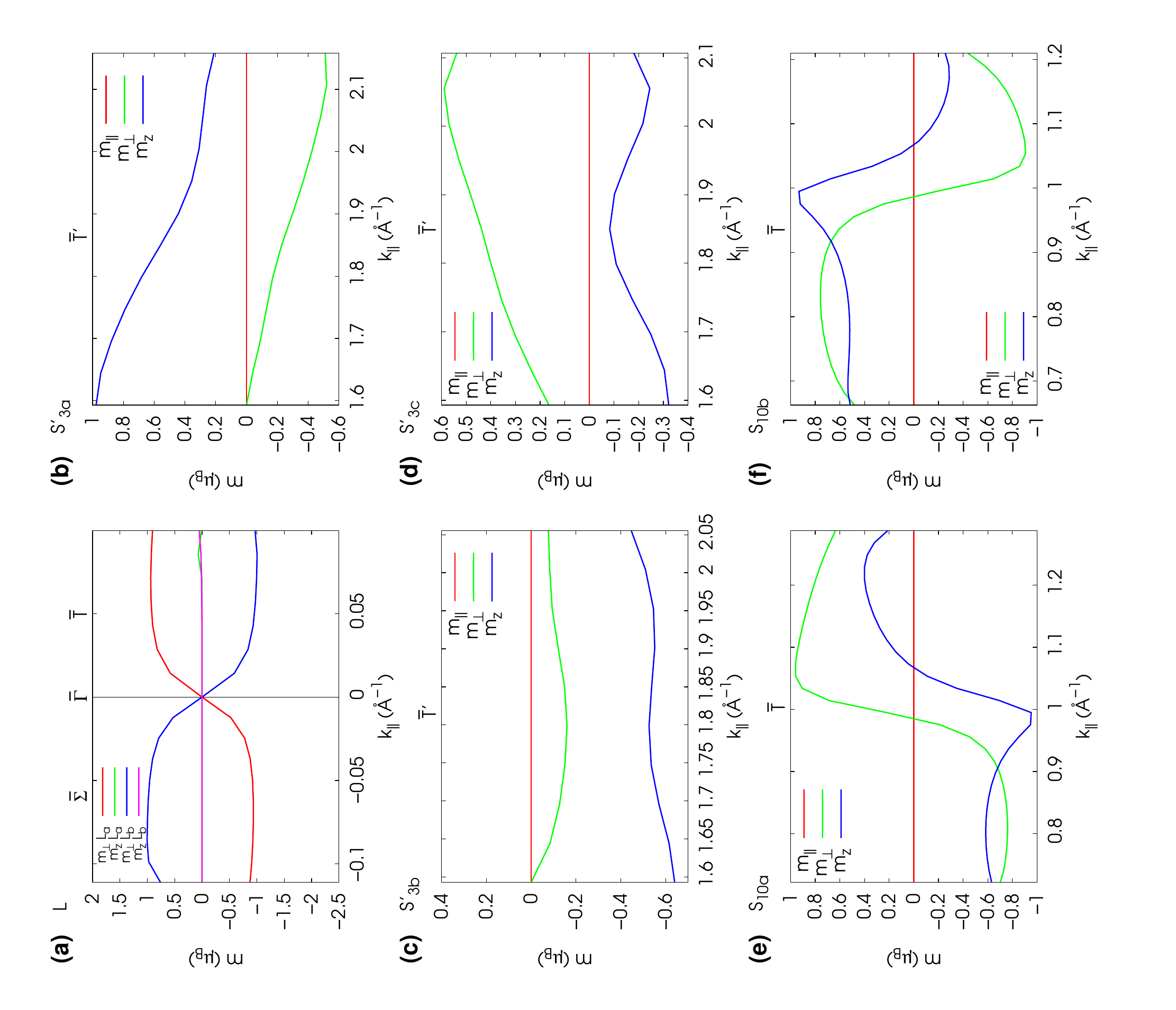}
\caption{Spin polarization components as a function of $\bm{k_{\parallel}}$ for the FR surface states $L_{a,b}$, $S'_{3a,b,c}$ and $S_{10a,b}$.}
\label{f7}
\end{figure}

\begin{figure}[h!]
\centering
\includegraphics[height=0.8\textwidth,angle=-90]{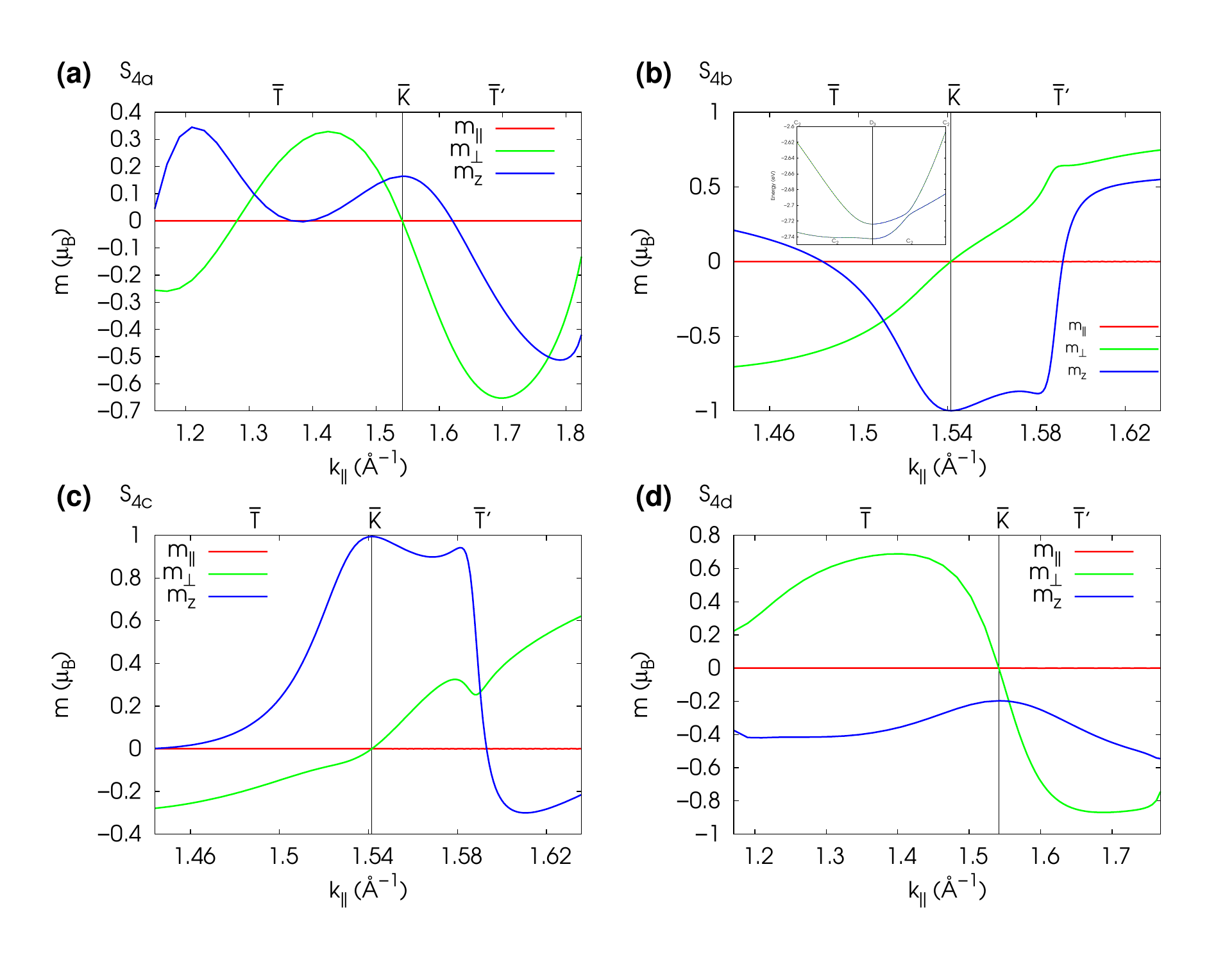}
\caption{Spin polarization components as a function of $\bm{k_{\parallel}}$ for the FR surface states $S_{4a,b,c,d}$.}
\label{f8}
\end{figure}

Along $\bar{T}$ and $\bar{T}'$ the spin polarization can rotate in a plane perpendicular to the high symmetry line and we investigate 
its rotation for a few surface states: $L$, $S'_{3}$, $S_{10}$, shown in Fig. \ref{f7}, and $S_4$, depicted in Fig. \ref{f8}. 
We start our analysis from the $L$ states (Fig. \ref{f7} a). Their spin polarization is mainly parallel to the surface and perpendicular to the high 
symmetry line, as predicted by the Rashba model, although a small component perpendicular to the surface survives along $\bar{T}$. 
The spin gets reverted when crossing the
$\bar{\Gamma}$ point, due to the different orientation of the $\bar{M}-\bar{\Gamma}$ and $\bar{\Gamma}-\bar{K}$ lines. 
The spin flipping is not sudden as predicted by the Rashba model, due to the
residue coupling between the two surfaces that opens a small gap at $\bar{\Gamma}$ (Fig. \ref{f9}).

Next we consider the $S'_3$ states (Fig. \ref{f7} (b-d)), named $S'_{3a}$, $S'_{3b}$, and $S'_{3c}$ in decreasing order of energy.
They show a quite smooth behavior, with a slowly varying spin polarization along the $\bar{T}'$
line, although there are some differences among them. The spin of $S'_{3a}$ state points outside the slab, the spins of $S'_{3b}$ and 
$S'_{3c}$ point towards the slab. Moreover, $S'_{3b}$ shows a small rotation of the spin, which points mainly along $z$, whereas in 
$S'_{3a}$ and $S'_{3c}$ the rotation is more evident: the spin is oriented mainly along $z$ at $\bar{K}$ and ends up
with a main component perpendicular to the high symmetry line.
 
The spin polarization of the $S_{10}$ states (Fig. \ref{f7}(e-f)) evolves more rapidly than in the 
$S'_3$ states: in particular, the most rapid variations are found at about $k_{\parallel} = 1 \, \AA^{-1}$.
The variations are due to the anticrossing of the two states (see Fig. \ref{f2}). 

Finally we analyze the spin polarizations of the $S_4$ states (Fig. \ref{f8}), that vary rapidly with $k_{\parallel}$
and show a quite complex behavior. Due to symmetry, the $m_{\perp}$ component must vanish at $\bar{K}$, whereas $m_z$ 
is not influenced by symmetry. The most regular evolution is shown by $S_{4d}$, for which the $m_z$ component
is almost constant and always negative; instead the other states show more evident rotations of the spin, which spans a 
wide range of different configurations in a quite small region of the high symmetry line. In particular, the $S_{4b}$ and $S_{4c}$ states 
show a quite abrupt variation of $m_z$ around $k_{\parallel} = 1.59 \, \AA^{-1}$. As for the $S_{10}$ states, this behavior can be explained
observing that at $k_{\parallel} \approx 1.59 \, \AA^{-1}$ $S_{4b}$ and $S_{4c}$ anti-cross. A more 
clear representation of the anti-crossing is given in the inset of Fig. \ref{f8}b, in which we show a magnification of the $S_{4b}$ and $S_{4c}$ 
states around the $\bar{K}$ point.

\section{Conclusions}
We presented a FR PAW LDA calculation of the electronic surface states of clean Os(0001). We described the nature and localization of the main surface
states and resonances analyzing their charge density contour levels and planar averages. 

We found a gap in the PBS around $\bar{\Gamma}$ similar to the L-gap typical of the (111) surfaces of the fcc lattice. Like in Ir(111) and at variance with Au(111) 
this gap does not host any surface state. We found Rashba split states close to $\bar{\Gamma}$, located below the L-gap. They cross the Fermi level, are 
hybridized with bulk states, and show a downward energy dispersion, as in Ir(111). The dispersion has been fitted with 
two Rashba-split parabolas with parameters $\gamma_{SO} = (6.1 \pm 0.1) \times 10^{-9} $ eV cm and $m^*/m= -0.280 \pm 0.002$.
We found $S_2$, $S'_{3}$, $S_4$, $S_6$, $S_7$, $S_{10}$, $S_{11}$, and $S_{12}$ states present also in the other surfaces, although $S_7$, $S_{11}$, and $S_{12}$ 
differ somewhat from the states found in Ir(111). The states $S_{1a}$, $S_{1b}$, and $S_9$ are not found with the charge density threshold 
used to identify the surface states in this work, but can be seen lowering this threshold. The $S_3$ states and  
the $S_8$ Dirac-like states, studied in Ref. \cite{dirac}, instead have not been found. We found
also a new group of states, that we named $S_{13}$, that appeared in the relaxed structure.

Moreover we analyzed the magnetization density of some selected surface states and computed their spin polarization.
We studied the spin polarization as a function of $\bm{k_{\parallel}}$ for some surface states along $\bar{T}$ and $\bar{T}'$, where
it can rotate in a plane perpendicular to the high symmetry line. 
It can vary from being smooth and slowly-varying to more complex with rapid rotations, which
we attributed to the mixing of the states.

The present work has been developed within the DFT-LDA scheme. The Kohn-Sham eigenvalues are distinct from the quasi-particle
energies, so in principle many-body corrections might be necessary for a detailed comparison with experiment. Yet, these calculations
are more computationally demanding and are usually carried out
only in those cases in which LDA is not sufficient to explain the experimental results. In the other surfaces, the main features of the
bands, such as the presence or absence of L-gap states, are well predicted by DFT-LDA, while the exact energy positions of the surface 
states might have small shifts. Unfortunately, to the best of our knowledge, no experimental measurement of the surface states is available 
for Os(0001), so we hope that our work could be of help for the experimental investigation of this surface and possibly, in case of discrepancies,
can motivate other theoretical calculations.

\section*{Acknowledgments}

Computational facilities have been provided by SISSA through its Linux Cluster and ITCS.
The work has been partially supported by the European Union
H2020-EINFRA-2015-1 programme under grant agreement No. 676598 project ``MaX - materials at the exascale''.







\begin{thebibliography}{00}
\bibitem{40_layers} A calculation with a 40-layer slab has been performed as well. The fit gives $\gamma_{SO} = (6.5 \pm 0.3) \times 10^{-9} $ eV cm, and $m^*/m= -0.132 \pm 0.002$.
\end{thebibliography}



\end{document}